\documentstyle[12pt]{article}
\begin{document}

\begin{center}
{\large\bf Transformation of the Extended Gamma Function
$\Gamma^{2,0}_{0,2}[(b,x)]$ with Applications to Astrophysical
Thermonuclear Functions}
        \\[1cm]
        M. Aslam Chaudhry\\
Department of Mathematical Sciences\\
King Fahd University of Petroleum and Minerals\\
Dhahran 31261, Saudi Arabia
\end{center}

\begin{abstract}
Two representations of the extended gamma functions
$\Gamma^{2,0}_{0,2}[(b,x)]$
are proved.  These representations are exploited to find a
transformation relation between two Fox's $H$-functions.  These
results are used to solve Fox's $H$-function in terms of Meijer's
$G$-function for certain values of the parameters.  A closed form
representation of the kernel of the Bessel type integral transform is
also proved.
\end{abstract}

\vspace{1cm}

\baselineskip=26pt
\noindent {\bf 1. \, Introduction}

\vspace{2mm}

\renewcommand{\thesection}{1.}
\renewcommand{\theequation}{\thesection\arabic{equation}}
\setcounter{equation}{0}
According to Anderson et al.(1994),
the
stars are gravitationally stabilized fusion reactors changing their
chemical composition while transforming light atomic nuclei into
heavy
ones.  The atomic nuclei are supposed to be in thermal equilibrium
with
the ambient plasma.  The majority of reactions among nuclei leading
to
a nuclear transformation are inhibited by the necessity for the
charged
particles to tunnel through their mutual Coulomb barrier.  The
theoretical and experimental verification of nuclear cross-sections
leads
to the derivation of the closed-form representation of
thermonuclear
reaction rates (Critchfield, 1972; Haubold and John, 1978; Haubold
and Mathai, 1986).
 These rates
are
expressed in terms of the four astrophysical thermonuclear
functions (Anderson et al., 1994)
\begin{eqnarray}
I_1 (z,\nu):= \int^\infty_0 y^
\nu \exp(-y-z/\sqrt{y})dy, \\
I_2 (z,d,\nu):= \int^d_0 y^
\nu \exp(-y-z/\sqrt{y})dy, \\
I_3 (z,t,\nu):= \int^\infty_0 y^
\nu \exp(-y-z/\sqrt{y+t})dy, \\
I_4 (z,\delta,b,\nu):= \int^\infty_0 y^
\nu \exp\left(-y-by^\delta - z/\sqrt{y}\right)dy. \end{eqnarray}
The closed-form representations of these integral functions,
asymptotic values and numerical results are discussed in Anderson
et al. (1994).

\vspace{2mm}

The closed-form representation of these functions in terms of the
Meijer's
$G$-function is essentially based on the following theorem
(Saxena, 1960;
Mathai and Haubold, 1988).  For $z > 0, p > 0, \rho \leq
0$,
and integers $m,n\geq 1$,
\begin{eqnarray}
&&  p \int^\infty_0 t^{-n\rho} \exp\left(-pt-zt^{-n/m}\right)dt \nonumber\\
&& \hspace*{5mm} \hspace*{5mm} = \; p^{n\rho}(2\pi)^{(2-n-m)/2}m^{1/2}
n^{\left(\frac{1}{2}\right)-n\rho} \times \nonumber\\
&&
\hspace*{5mm} \hspace*{5mm} \hspace*{5mm}
G^{m+n,0}_{m+n,0}
\left(
\frac{z^m p^n}{m^n n^n} \left| \begin{array}{c}
-,\; \; -, \ldots, \;\; - \\
0, \frac{1}{m}, \frac{2}{m},
\ldots, \frac{m-1}{m}; \;\; 1- \frac{n\rho}{n}, \; \; n - \frac{n\rho}
{n}\end{array} 
\right.\right).
\end{eqnarray}
The restriction $\rho \leq 0$ is essential in the proof of (1.5)
whereas one would like to have a closed-form representation free of
this restriction.  This has been achieved in this paper.

\vspace{2mm}

The study of the astrophysical thermonuclear functions led   to the
development of a new class of special functions (Chaudhry and
Zubair, 1998) that
extends Meijer's $G$ and Fox's $H$-function.  In this paper we
prove a transformation relation for the extended gamma functions
(Chaudhry and Zubair, 1998)
\begin{equation}
\Gamma(\alpha,x;b,\beta):= \int^\infty_x t^{\alpha-1} e^{-t-b/t^\beta}dt,
\end{equation}
and
\begin{equation}
\gamma(\alpha,x;b,\beta):= \int^x_0 t^{\alpha-1} e^{-t-b/t^\beta}dt,
\end{equation}
and derive the closed-form of the astrophysical thermonuclear
functions
in terms of Meijer's $G$-function.  It is to be noted that the
functions (1.6) and (1.7) are special cases of the general class of
extended gamma functions introduced in Chaudhry and Zubair (1998).

In
fact we have
\begin{equation}
\Gamma(\alpha,x;b,\beta)= \Gamma^{2,0}_{0,2} \left[(b,x)\left|
\begin{array}{cc}
-, & - \\
((0,1), & (\alpha,\beta)) \end{array} \right.\!\!\right] := \frac{1}{2\pi i}
\int^{C+i\infty}_{C-i\infty} \Gamma(s)\Gamma(\alpha+\beta s,x)b^{-s}ds,
\end{equation}
and
\begin{equation}
\gamma(\alpha,x;b,\beta)= \gamma^{2,0}_{0,2} \left[(b,x)\left|
\begin{array}{cc}
-, & - \\
((0,1), & (\alpha,\beta)) \end{array} \right.\!\!\right] := \frac{1}{2\pi i}
\int^{C+i\infty}_{C-i\infty} \Gamma(s)\gamma(\alpha+\beta s,x)b^{-s}ds.
\end{equation}
It is important to note that the astrophysical thermonuclear
functions (1.1) and (1.4) are special cases of the extended
gamma functions (1.8) and (1.9).  As a matter of fact we have
\begin{eqnarray}
I_1 (z,\nu) & = & \Gamma\left(\nu + 1, 0; z; \frac{1}{2}\right),\\
I_2 (z,d,\nu) & = & \gamma\left(\nu + 1, d; z; \frac{1}{2}\right),\\
I_3 (z,t,\nu) & = & e^t \sum^\nu_{r=0} \left(\begin{array}
{c} \nu \\ r \end{array} \right)
(-t)^{\nu-r}
\Gamma\left(\nu + 1, t; z; \frac{1}{2}\right),\\
I_4 (z,\delta,b,\nu) & = &   \sum^\infty_{r=0} \frac{(-b)^r}{r!}
\Gamma\left(\nu + r \delta+1, 0; z; \frac{1}{2}\right).
\end{eqnarray}

\vspace{8mm}

\noindent {\bf 2. The Transformation Theorem}

\renewcommand{\thesection}{2.}
\renewcommand{\theequation}{\thesection\arabic{equation}}
\setcounter{equation}{0}
\vspace{6mm}

\noindent {\bf Theorem (2.1)}. {\em For $x \geq 0, b\geq 0, \beta \geq 0$,
\begin{equation}
\Gamma(\alpha,x;b,\beta) = \frac{1}{\beta} \Gamma^{2,0}_{0,2}
\left[\left(b^{1/\beta},x\right)\left|
\begin{array}{cc}
- & - \\
\left(\left(0, \frac{1}{\beta}\right)\right., & \left.
\left(\alpha,1\right)\right)\end{array} \right.\!\!\right]
\end{equation} }

\noindent {\bf Proof}. Replacing $b$ by $b^\beta$ in (1.6) yields
\begin{equation}
\Gamma(\alpha,x;b^\beta;\beta) = \int^\infty_0 f_1(b/t)f_2(t)\frac{dt}{t},
\end{equation}
where
\begin{eqnarray}
f_1(t):= e^{-t^\beta},\\
f_2(t):= t^\alpha e^{-t}H(t-x), \end{eqnarray}
and
\begin{equation}
 H(t-x):= \left\{\begin{array}{ll}
 1 & \hspace*{5mm} \mbox{ if } \; t > x,\\
 0 & \hspace*{5mm} \mbox{ if } \; t < x. \end{array} \right.\,
 \end{equation}
The Mellin transforms of the functions (2.3) and (2.4) are readily
found
to be
\begin{eqnarray}
\overline{f}_1(s):= \frac{1}{\beta}\Gamma(s/\beta)\\
\overline{f}_2(s):= \Gamma(\alpha+s,x). \end{eqnarray}
However, according to Erd\'{e}lyi et al. (1954)
\begin{equation}
M\left(\int^\infty_0 f_1(b/t)f_2(t)\frac{dt}{t}; s \right)
=
\overline{f}_1(s) \overline{f}_2(s).
\end{equation}
From (2.2), (2.6) and (2.7) we have
\begin{eqnarray}
\Gamma(\alpha,x;b^\beta;\beta) & = & \frac{1}{\beta}
M^{-1}\left\{\Gamma\left(\frac{s}{\beta}\right)\Gamma
(\alpha+s,x);b\right\} \\
& = & \frac{1}{\beta} \frac{1}{2\pi i} \int^{C+i\infty}_{C-i\infty}
\Gamma\left(\frac{s}{\beta}\right) \Gamma(\alpha+s,x)b^{-s}ds\\
& = & \frac{1}{\beta} \Gamma^{2,0}_{0,2} \left[(b,x)\left|
\begin{array}{cc}
-, & - \\
\left(\left(0, \frac{1}{\beta}\right), \right. & \left. (\alpha,1)
\right) \end{array} \right.\!\!\right].
\end{eqnarray}
Replacing $b$ by $b^{1/\beta}$ in (2.11) yields (2.1).

\vspace{8mm}

\noindent {\bf 3. Applications of the Transformation to
Astrophysical Thermonuclear Functions}

\renewcommand{\thesection}{3.}
\renewcommand{\theequation}{\thesection\arabic{equation}}
\setcounter{equation}{0}
\vspace{6mm}

\noindent {\bf Theorem (3.1)}.
\begin{equation}
\Gamma(\alpha,0;b;\beta) = H^{2,0}_{0,2}
\left[
b \left| \begin{array}{cc}
-, & - \\
(0,1), & (\alpha,\beta)\end{array} \right.\!\! \right]
= \frac{1}{\beta} H^{2,0}_{0,2} \left[
b^{1/\beta} \left|\begin{array}{c} -, \;\; - \\
(\alpha,1), \;\; \left(0, \frac{1}{\beta}\right)\end
{array} \right.\!\!\right].
\end{equation}

\noindent {\bf Proof}. This is a direct consequence of (1.8) and (2.1).

\vspace{2mm}

\noindent {\bf Corollary (3.1)}.
\begin{eqnarray}
&&
\Gamma\left(\alpha,0;b;\frac{1}{n}\right) =
H^{2,0}_{0,2} \left[b\left|
\begin{array}{cc}
-, & - \\
(0,1), & (\alpha,n) \end{array} \right.\!\!\right] =
(2\pi)^{(1-n)/2}\sqrt{n} \times \nonumber\\
&& \hspace*{1cm}
G^{n+1,0}_{0,n+1} \left[
\left(\frac{b}{n}\right)^n \left|
\begin{array}{ccc}
\\ -, &  -, \ldots, & - \\
0, & \frac{1}{n}, \frac{2}{n}, \ldots, \frac{n-1}{n}, & \alpha \end
{array} \right.\!\!\right]
\end{eqnarray}

\noindent {\bf Proof}. The substitution $\beta = \frac{1}{n}$ in (3.1) leads
to
\begin{eqnarray}
\Gamma\left(\alpha,0;b;\frac{1}{n}\right) & = & n H^{2,0}_{0,2} \left[
b^n \left|
\begin{array}{cc}
-, & - \\
(\alpha,1), & (0,n)\end{array} \right.\!\!\right] \nonumber\\
& = & n \frac{1}{2\pi i} \int^{C+i\infty}_{C-i\infty} \Gamma(\alpha+s)
\Gamma(ns) b^{-ns}ds.
\end{eqnarray}
However, the use of the multiplication formula
(Anderson et al., 1994; Mathai and Haubold, 1988)
\begin{equation}
\Gamma(mz) = (2\pi)^{(1-m)/2} m^{mz- \frac{1}{2}}
\prod^{m-1}_{k=0} \Gamma\left(z + \frac{k}{m}\right)
\end{equation}
for the gamma function yields
\begin{equation}
\Gamma\left(\alpha,0;b;\frac{1}{n}\right) = (2\pi)^{(1-n)/2}\sqrt{n}
\frac{1}{2\pi i} \int^{C+i\infty}_{C-i\infty}
\Gamma(\alpha+s) \prod^{n-1}_{k=0}
\Gamma\left(s + \frac{k}{n}\right) (b/n)^{-ns}ds,
\end{equation}
which is exactly (3.2).

\vspace{2mm}

\noindent {\bf Corollary (3.2)}.
\begin{equation}
\Gamma\left(\alpha,0;b;\frac{1}{2}\right) = \pi^{-1/2} G^{3,0}_{0,3}
\left[\frac{b^2}{4}\left|_{0, \frac{1}{2}, \alpha}\right.\right].
\end{equation}

\noindent {\bf Proof}.
This is a special cases of (3.2) when we take $n=2$.

\vspace{2mm}

\noindent {\bf Remark}. The closed-form representation (3.6) is
important
in view of the relation (1.10) that yields (Anderson et al., 1994)
\begin{eqnarray}
I_1(z,\nu) & = & \Gamma\left(\nu + 1, 0;z; \frac{1}{2}\right) \nonumber\\
& = & \pi^{-1/2} G^{3,0}_{0,3} \left[
\frac{z^2}{4} \left|_{0, \frac{1}{2}, 1+ \nu}\right.\right].
\end{eqnarray}
Moreover,
\begin{equation}
I_2(z,d,\nu) =
\Gamma\left(\nu + 1, 0; z; \frac{1}{2}\right) - \Gamma \left(\nu + 1, d; z;
\frac{1}{2}\right),
\end{equation}
and the functions $I_3(z,t;\nu)$ and $I_4(z,\delta,b,\nu)$ are
expressible in terms of $I_1$ and $I_2$ functions.  Therefore, it
seems
important to search for the closed-form of the extended function
$\Gamma\left(\alpha,x;b; \frac{1}{2}\right)$ in terms of classical special
functions.  In view of the results proved in Anderson et al. (1994),
it seems impossible to have such type of
representations.
Thus, the extended gamma functions (Chaudhry and Zubair, 1998)
provide
the unique closed form-representation of the astrophysical
thermonuclear functions given by (1.10) -- (1.13).

\vspace{8mm}

\noindent {\bf 4. Application to Bessel Type Integral Transforms}

\vspace{6mm}

\renewcommand{\thesection}{4.}
\renewcommand{\theequation}{\thesection\arabic{equation}}
\setcounter{equation}{0}
Kilbas et al. (1998)
      have studied the integral transform
\begin{equation}
K^\rho_\nu(f)(x) = \int^\infty_0 z^\nu_\rho(xt)f(t)dt, \hspace*{5mm} (x > 0),
\end{equation}
with the kernel
\begin{equation}
z^\nu_\rho(x) = \int^\infty_0 t^{\nu-1}\exp\left(-t^\rho -
\frac{x}{t}\right)dt,
\hspace*{5mm} (\rho > 0, \nu \in {\rm I\hskip-6.5pt C}),
\end{equation}
on the spaces $F_{p,\mu}$ and $F'_{p,\mu}$ $(1 \leq p \leq \infty,
u\in {\rm I\hskip-6.5pt C})$ of tested and generalized functions. 
When $\rho =1$ and
$x = t^2/4$
\begin{equation}
z^\nu_1 (t^2/4) = 2(t/2)^\nu K_\nu(t),
\end{equation}
where $K_\nu(t)$ is the modified Bessel function of the third kind.
For other values of $\rho$, the authors  considered the
integral representation (4.2) of the kernel {\em without}
     the closed-form representation, and
            proved the compositions of the operator $K^\rho_\nu$
with
the left and right-sided Liouville fractional integrals and
derivatives.

\vspace{2mm}

The left-sided $I^\alpha_{0+}, D^\alpha_{0+}$ and right-sided $I^\alpha_-,
D^\alpha_-$ Liouville fractional derivatives are defined for $x > 0$
by (Kilbas et al., 1998)
\begin{eqnarray}
\left(I^\alpha_{0+} \varphi\right)(x) & = &
\frac{1}{\Gamma(\alpha)} \int^x_0 \frac{\varphi(t)dt}{(x-t)^{1-\alpha}},
\hspace*{5mm} (\alpha > 0),\\
\left(D^\alpha_{0+} \varphi\right) (x) & = & \left(
\frac{d}{dx}\right)^{[\alpha]+1}
\left(I^{1-\{\alpha\}}_{0+}
\varphi\left(x\right)\right), \hspace*{5mm} (\alpha > 0),\\
\left(I^\alpha_- \varphi\right)(x) & = & \frac{1}{\Gamma(\alpha)}
\int^\infty_x \frac{\varphi(t)dt}{(t-x)^{1-\alpha}}, \hspace*{5mm} 
(\alpha > 0), \\
\left(D^\alpha_-\varphi\right)(x) & = & \left(-
\frac{d}{dx}\right)^{[\alpha]+1}
\left(I^{1-\{\alpha\}}\varphi\right)(x), \hspace*{5mm} (\alpha > 0),
\end{eqnarray}
respectively, where $[\alpha]$ and $\{\alpha\}$ are the integral and
fractional parts of $\alpha > 0$.

\vspace{2mm}

The following relations are then
proved:
\begin{eqnarray}
\left(K^\rho_\nu I^\alpha_{0+}\right)\varphi & = &
\left(x^{-\alpha}K^\rho_{\nu+\alpha}\right)
\varphi,\\
\left(K^\rho_\nu D^\alpha_{0+}\right)\varphi & = & \left(x^{\alpha}
K^\rho_{\nu-\alpha}\right)
\varphi,\\
\left(I^\alpha_- D^\rho_\nu\right)\varphi & = &
K^\rho_{\nu+\alpha}\left(x^{-\alpha}\varphi\right),
\\
\left(D^\alpha_- K^\rho_\nu\right)\varphi & = &
K^\rho_{\nu-\alpha}\left(x^{\alpha}\varphi\right)
. \end{eqnarray}
It is to be noted that the kernel (4.2) of the integral transform
(4.1)
can be simplified in terms of the extended gamma function
\begin{equation}
\Gamma(\alpha,0;b;\beta) = H^{2,0}_{0,2} \left[b\left|
\begin{array}{cc}
-, & - \\
(0,1), & (\alpha,\beta) \end{array} \right.\!\!\right].
\end{equation}
The transformations
\begin{equation}
t^\rho = u, \hspace*{5mm} \frac{dt}{t} = \frac{1}{\rho} \frac{du}{u}
\end{equation}
in (4.2) lead  to
\begin{equation}
z^\nu_\rho (x) = \frac{1}{\rho}
\int^\infty_0 u^{\nu/\rho-1} \exp\left(-u-
\frac{x}{u^{1/\rho}}\right)du,
\end{equation}
which is expressible in terms of the
extended gamma function to give (Chaudhry and Zubair, 1998)
\begin{equation}
z^\nu_\rho(x)   =   \frac{1}{\rho} \Gamma\left(\frac{\nu}{\rho}, 0; x;
\frac{1}{\rho}\right), \end{equation}
which can further be simplified in terms of the Fox's $H$-function
to give
\begin{equation}
z^\nu_\rho(x)
  =   \frac{1}{\rho} H^{2,0}_{0,2} \left[x \left|
\begin{array}{cc}
-, & - \\
(0,1), & \left(\frac{\nu}{\rho}, \frac{1}{\rho}\right) \end
{array} \right.\!\!\right].
\end{equation}
Hence the transformation (4.1) can be defined in a closed-form as
follows:
\begin{equation}
\left(K^\nu_\rho f\right)(x) = \frac{1}{\rho} \int^\infty_0 H^{2,0}_{0,2}
\left[
xt \left|\begin{array}{cc}
-, & - \\
(0,1), & \left(\frac{\nu}{\rho}, \frac{1}{\rho}\right)\end
{array} \right.\!\!\right]
f(t)dt.
\end{equation}
The representation (4.15) -- (4.17)  and the relations (4.7) -- (4.10) can be
exploited to find the closed-form representations of the integrals
involving the extended gamma function (4.11).  In particular, the
substitution $\varphi(x) = \delta(x)$ in (4.8) yields
\begin{equation}
\int^\infty_0 t^{\alpha-1} H^{2,0}_{0,2} \left[
xt \left|
\begin{array}{cc}
-, & - \\
(0,1), & \left( \frac{\nu}{\rho},
\frac{1}{\rho}\right) \end{array} \right.\!\!\right]dt =
\Gamma(\alpha) \Gamma \left( \frac{\nu + \alpha}{\rho}\right)x^{-\alpha},
\hspace*{5mm} (\alpha > 0).
\end{equation}
The closed form representation of the kernel (4.2) in terms
of Fox's $H$-function has provided a compact and useful
representation
(4.17) of the integral transformation (4.1).  It will facilitate
the
study of the integrals involving astrophysical thermonuclear
functions.

\vspace{8mm}

\noindent {\bf Acknowledgments}. The author is indebted to the King
Fahd
University of Petroleum and Minerals for providing excellent
research
facilities.  Private communications with Professor H.J. Haubold are
appreciated.

\newpage

\begin{center} {\bf References} \end{center}

\baselineskip=13.5pt
\noindent Anderson, W.J., Haubold, H.J., and Mathai, A.M.
(1994).
Astrophysical thermonuclear functions. {\em Astrophys. Space
Sci.}, {\bf 214}, 49--70\\ 
(http://xxx.lanl.gov/abs/astro-ph/9402020). \\[4mm]
\noindent Chaudhry, M.A. and Zubair, S.M. (1998). Extended
incomplete gamma functions with applications, {\em Journal
of London Mathematical Society} (submitted). \\[4mm]
\noindent Critchfield, C.L. (1972). In: Cosmology, Fusion
and Other Matters, George Gamow Memorial Volume. Ed. F.
Reines, Colorado, Colorado: Associated University Press 1972.
\\[4mm]
\noindent Erd\'{e}lyi et al. (1954). {\em Table of Integral
Transforms}, Volume 1, McGraw-Hill, New York.
\\[4mm]
\noindent Haubold, H.J. and Mathai, A.M. (1986). Analytic
representations of thermonuclear reaction rates. {\em Studies
in Applied Mathematics}., {\bf LXXV  (2)}, 123--137. \\[4mm]
\noindent Haubold, H.J., John, R.W. (1978). On the evaluation
of an integral connected with the thermonuclear reaction rate
in closed form. {\em Astron. Nachr}. {\bf 299}, 225--232; {\bf 300},
173. \\[4mm]
\noindent Haubold, H.J. and Mathai, A.M. (1986). Analytic results
for screened non-resonant nuclear reaction rates, {\em Astrophys.
Space Sci.}, {\bf 127}, 45--53. \\[4mm]
\noindent Kilbas, A.A., Bonilla, B., Rivero, M., Rodrigues, J.
and
Trujillo, J. (1998). Composition of Bessel type integral transform
with fractional operators on spaces $F_{p,\mu}$ and $F'_{p,\mu}$\,,
{\em Fractional Calculus and Applied Analysis}, {\bf 1}, no. 2,
135.
\\[4mm]
Mathai, A.M. and Haubold, H.J. (1988). {\em Modern Problems
in Nuclear and Neutrino Astrophysics}, Akademie-Verlag, Berlin. \\[4mm]
Saxena, R.K. (1960). {\em Proc. Nat. Acad. Sci. India}, {\bf 26},
400--413.
\\[4mm]
\end{document}